# Channeling as a method of making nanosized beams of particles

V.M. Biryukov[•]

*Institute for High Energy Physics, Protvino, RU-142281, Russia*

**Abstract**

Particle channeling in a bent crystal lattice has led to an efficient instrument for beam steering at accelerators [1], demonstrated from MeV to TeV energies. In particular, crystal focusing of high energy protons to micron size has been demonstrated at IHEP with the results well in match with Lindhard (critical angle) prediction. Channeling in crystal microstructures has been proposed as a unique source of a microbeam of high-energy particles [2]. Channeling in nanostructures (single-wall and multi-wall nanotubes) offers the opportunities to produce ion beams on nanoscale. Particles channeled in a nanotube (with typical diameter of about 1 nanometer) are trapped in two dimensions and can be steered (deflected, focused) with the efficiency similar to that of crystal channeling or better. This technique has been a subject of computer simulations, with experimental efforts under way in several high energy labs, including IHEP. We present the theoretical outlook for making channeling-based nanoscale ion beams and report the experience with crystal-focused microscale proton beams.

*Keywords:* Channeling, nanotube, microbeam, nanobeam.

*PACS*: 61.85.+p; 02.40.-k

## 1. Introduction

In the past decade, the technique of particle beam channeling by bent crystals has well progressed into an established beam instrument at GeV accelerators [1]. Remarkably, it was demonstrated in the energy range spanning over six decades [3-15], from 3 MeV [6,7] to 900 GeV [8-10]. IHEP Protvino has pioneered the wide practical use of bent crystals as optical elements in high-energy beams for beam extraction and

---
[•] E-mail: Valery.Biryukov@ihep.ru

deflection on permanent basis since 1989. In the course of IHEP experiments, crystal channeling has been developed into efficient instrument for particle steering at accelerators, working in predictable, reliable manner with beams of very high intensity over years [11-13,16,17]. Crystal systems extract 70 GeV protons from IHEP main ring with efficiency of 85% at intensity of $10^{12}$ proton per spill (of about 1 s duration), steered by silicon crystal of just 2 mm in length [12]. One of the crystals was exposed for several minutes to even higher radiation flux of 70 GeV protons, about ~$10^{14}$ proton hits per spill of 50 ms with repetition period of 9.6 s. After the exposure, the channeling properties of the crystal were found normal as shows photo in Fig. 1.

Today, six locations on the IHEP 70-GeV main ring of the accelerator facility are equipped by crystal extraction systems, serving mostly for routine applications rather than for research and allowing a simultaneous run of several particle physics experiments, thus significantly enriching the IHEP physics program. One crystal served in the vacuum chamber of the IHEP U-70 over 10 years, from 1989 to 1999, delivering beam to particle physicists, until a new crystal replaced it. The long successful history of large-scale crystal exploitation at IHEP should help to incorporate channeling crystals into accelerator systems worldwide in order to create unique systems for beam creation and delivery.

## 2. Crystal focusing experiment

Following the successful application of bent crystals as miniature "dipoles" for beam bending, an experiment started at IHEP aiming to show that crystal could also focus a particle beam. The idea of crystal focusing [18-21] is explained by Fig. 2. The deflection angle any particle obtains in the crystal should be a function of particle's transverse coordinate so as to make all channeled particles cross the focus some distance downstream of the crystal. A series of focusing crystals was made with focus lengths F from 0.5 to 3.5 m and tested in 70 GeV proton beam [19-21].

Fig. 3 shows the image at the crossover of the beam focused (and deflected) by the crystal with F=0.5 m. The profile of the deflected and focused beam can be seen on the left. The dashed rectangle on the right is the crystal cross-section. In this example, the measured beam size in the focus is just 21 micron r.m.s. Table 1 provides the measured data for three tested crystals. It shows also the expected beam size given by the divergence of channeled particles (Lindhard angle) times the focus length. Remarkably, for two of the crystals the predicted and measured data are well in match; for the third crystal, with the shortest focus, the aberration in size is seen.

## 3. Nanotube channeling

There has been large interest recently [22-26] to possibility of channeling in the materials produced by nanotechnology. Carbon nanotubes are cylindrical molecules with a typical diameter of order 1 nm for single-wall nanotubes (SWNT) or a few tens nm for multi-wall nanotubes (MWNT) and a length of many microns [27]. They are made of carbon atoms and can be thought of as a graphene sheet rolled around a cylinder, Fig.4. Creating efficient channeling structures - from single crystals to nanotubes - might have a significant effect onto the accelerator world. Ideally, we would like to trap particles in two dimensions; to have the channel walls made of densely packed atoms; to make channel size of our choice; to build it with atoms of our choice. This is indeed close to what we get from nanotechnology nowadays.

The potential within a nanotube is localised very close to the wall, Fig. 5 (a). Essential for steering of particles by nanotube is how the potential well modifies by centrifugal term with bending of nanotube. Fig. 5 (b) reveals that much of the nanotube cross-section is still available for channeling even with bending of $pv/R$=1 GeV/cm (equivalent to 300 Tesla), $R$ being the bending radius, $pv$ particle momentum times velocity. For understanding of bending dechanneling in nanotubes of different size and geometry, we did Monte Carlo simulations of particle channeling in bent single-wall and multi-wall nanotubes, aiming to find how useful are the nanotubes for channeling of positively-charged particle beams, what kind of nanotubes are efficient for this job, and how nanotubes compare with crystals in this regard. The particles were tracked through the curved nanotubes in the approach with a continuous potential described in more detail elsewhere [28]. Fig. 6 shows the number of the channeled protons as a function of the nanotube curvature $pv/R$ for MWNT and SWNT (0.4-nm diameter), and for Si (110) crystal. This comparison shows that steering capabilities of different channeling structures are similar, for the range of parameters shown in the Figure.

## 4. Emittance of nanobeam

Two approaches to making a small beam by means of channeling can be pursued. One can trap a small fraction of the incident beam and steer it away in order to form a low-emittance beam with well-defined sharp edges that contains solely primary particles. Another possibility would be to arrange a focusing array of bent channeling nanotubes and focus (in two dimensions) the channeled particles into a small spot; this focusing approach is certainly much more challenging technologically, therefore we restrict the discussion below to the first approach. The possibility to deflect the trapped nanobeam out of the direction of primary beam is a clear advantage over a passive capillary collimation used in microbeam facilities.

The size of a channeled nanobeam is set by the size of channeling nanostructure. The diameter of a typical SWNT is about 1 nm (the narrowest one is about 0.4 nm); the diameter of a rope of SWNTs is usually about 30 nm. The external diameter of MWNT is 30-50 nm.

The depth $U_0$ of the potential well determines the angular divergence of the channeled beam. The critical angle for channeling is about $\theta_c=(2U_0/pv)^{1/2}$. In a carbon nanotube with an arbitrary helicity the channeled particles are confined in a potential well with $U_0 \approx 60$ eV. This figure doesn't depend practically on the size of the channel, as the atomic field within the nanotube falls off sharply in a close vicinity of the wall already. Therefore, the emittance of the nanobeam formed by a single channeling SWNT could start from about 0.001 π nm·radian in a 100 MeV range, or 0.1 π nm·radian in a 1 MeV range.

With small emittance, the intensity of nanobeam would also be small. The effects of thermal shock and radiation damage, eventually reducing the structure lifetime, may restrict the intensity also. The IHEP experience shows that crystals can channel up to ~3·10$^{12}$ particles/s per cross-section of 0.5×5 mm$^2$ without cooling measures. For nanostructures there is not enough data, although damage processes are being evaluated [29]. However crystals show a huge margin of safety in lattice lifetime and thus raise hope for similar stability of nanostructures. The above figure for crystal survival corresponds to the intensity of $10^6/(s \cdot \mu m^2) = 1/(s \cdot nm^2)$. Extrapolating, we would expect a 1-nm-sized SWNT to channel 1 proton/s without problem, or a 30-nm-sized nano-rope or MWNT to channel 1000 proton/s. This is interesting enough; the applications like microbeam facilities often require quite small intensities, down to 1-1000 particles/s. A lifetime of $\sim 5 \cdot 10^{20}$ proton irradiation per $cm^2$ was measured [5] for channeling crystal. This corresponds to $5 \cdot 10^6/nm^2$ and means almost a year of continuous operation of nanobeam at 1 proton per $nm^2$ per second.

## 5. Conclusion

Crystal channeling is a well-developed beam instrument nowadays. The nanotube channeling is already well understood by theory. From physics standpoint, it is possible to make nano-beams by means of channeling. However, this is challenging task for technology. With strong user demand, this field may grow indeed from academic research to application.

**Acknowledgements.**

This work was partially supported by INFN - Gruppo V, as NANO experiment, and by INTAS-CERN grants 132-2000 and 03-52-6155.

**Table 1** Beam size in focus for different crystals.

| Crystal N | F (m) | R.m.s. size (μm), measured | Size (μm), calculated |
|---|---|---|---|
| 1 | 3.5 | 100 | 88 |
| 2 | 1.4 | 40 | 35 |
| 3 | 0.5 | 21 | 12 |

# Figure captions

**Figure 1**

Photograph of the deflected (left) and incident (right) beams as seen downstream of the crystal. Prior to the test, the crystal was exposed in the ring to 50-ms pulses of very intense beam (~$10^{14}$ proton hits per pulse). No damage of crystal was seen in the test, after this extreme exposure.

**Figure 2**

Principle of beam focusing by a crystal. $II'$ is the focal line where the tangents to the bent planes converge.

**Figure 3**

The image at the crossover of the beam focused (and deflected) by a crystal. The profile of the deflected and focused beam can be seen on the left. The dashed rectangle on the right is the crystal cross-section.

**Figure 4**

  (a) Schematic view of SWNT.
  (b) STM image (900nm x 900 nm) of SWNT ropes.
  (c) Schematic view of MWNT.
  (d) TEM image of a MWNT (about 20 walls) shown on a scale of 5 nm.

**Figure 5**

  (a) The continuous potential within a carbon nanotube (arbitrary helicity).
  (b) The effective potential within a nanotube bent with $pv/R=1$ GeV/cm.

**Figure 6**

The number of the channeled protons as a function of the nanotube curvature $pv/R$ for MWNT and SWNT (0.4-nm diameter), and for Si (110) crystal.

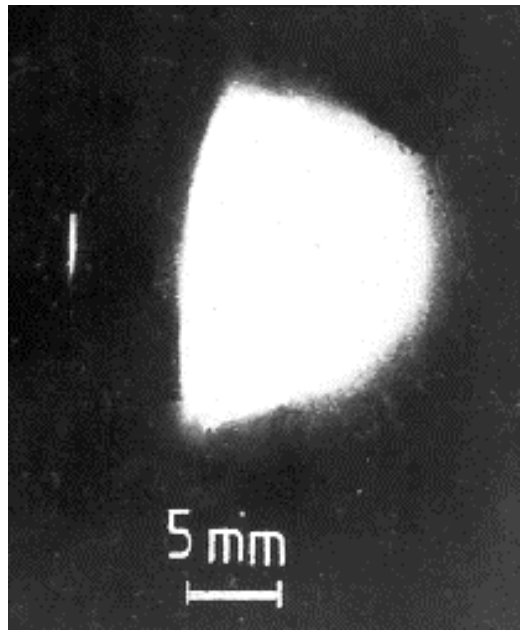

**Figure 1**

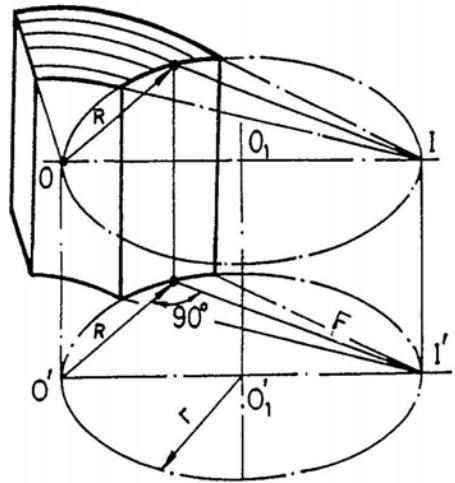

**Figure 2**

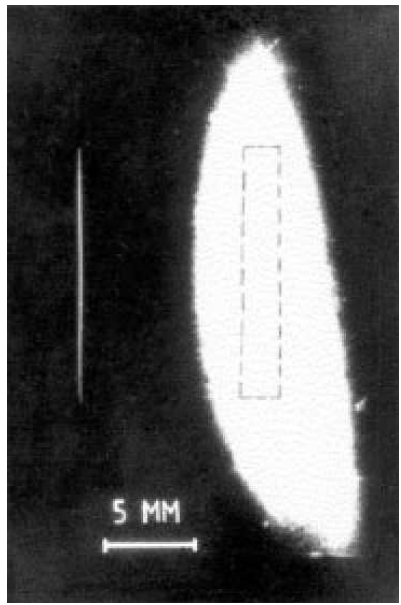

**Figure 3**

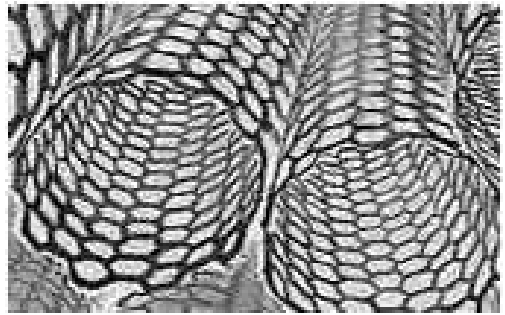

Figure 4 (a)

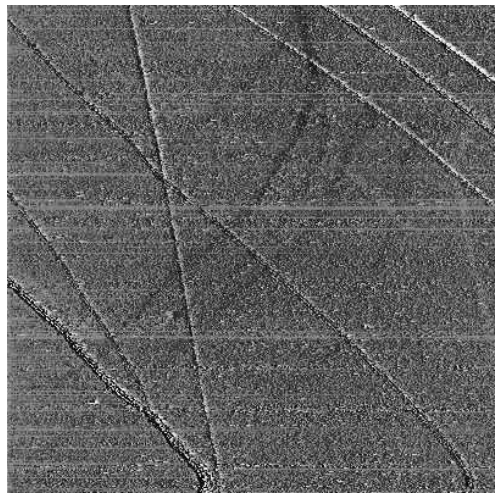

Figure 4 (b)

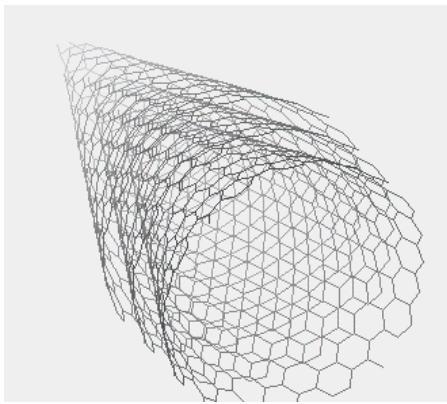

Figure 4 (c)

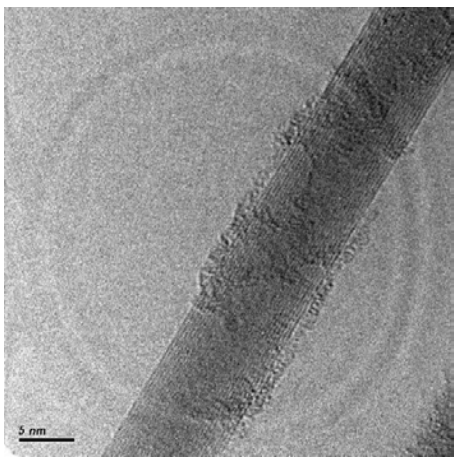

Figure 4 (d)

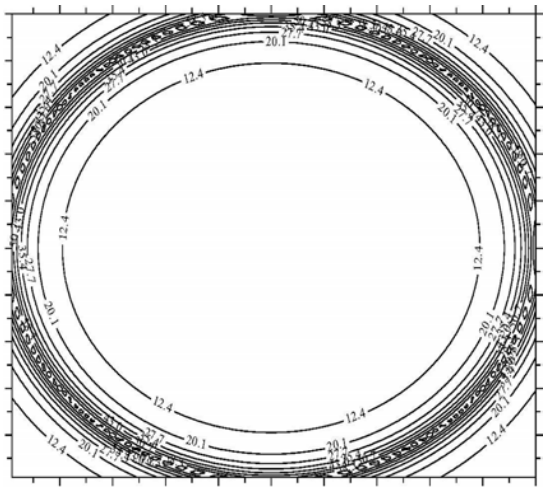

Figure 5 (a)

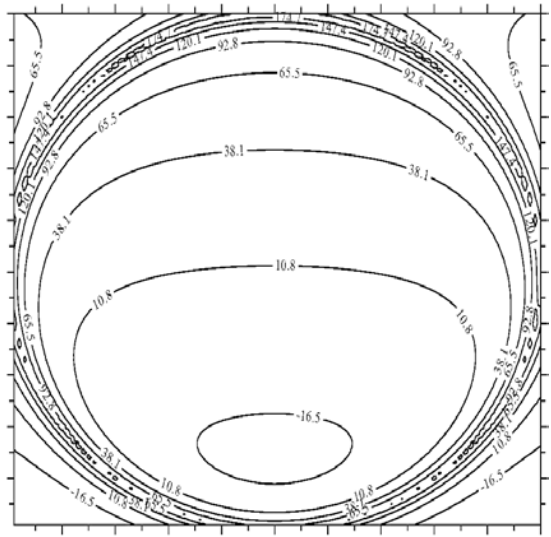

Figure 5 (b)

Figure 6